\documentclass[apj,numberedappendix]{emulateapj}
\usepackage{amsmath}
\usepackage{txfonts}
\usepackage{graphicx}

\newcommand{\notea}{\rlap{\textsuperscript{a}}}

\def\p0{\phantom{0}}

\def\***#1{\textsf{***#1***}}

\begin{document}

\shorttitle{MASS-TEMPERATURE RELATION} \shortauthors{KOTOV \& VIKHLININ}
\slugcomment{To be submitted to ApJ Letters} 

\title{{\em Chandra} SAMPLE of GALAXY CLUSTERS at $z=0.4-0.55$:\\
  EVOLUTION IN THE MASS-TEMPERATURE RELATION} 

\author{O.  Kotov\altaffilmark{1}\altaffilmark{,2} and A.
  Vikhlinin\altaffilmark{2}\altaffilmark{,1}} \altaffiltext{1}{Space
  Research Institute, Moscow, Russia; kotov@head.cfa.harvard.edu}
\altaffiltext{2}{Harvard-Smithsonian Center for Astrophysics, 60 Garden
  St., Cambridge, MA 02138}

\begin{abstract}
  We present spatially-resolved analysis of the temperature and gas
  density profiles in 6 relaxed galaxy clusters at $z = 0.4-0.54$ using
  long-exposure \emph{Chandra} observations. We derived the total
  cluster masses within the radius $r_{500}$ assuming hydrostatic
  equilibrium but without assuming isothermality of the intracluster
  gas. Together with a similar study based on the \emph{XMM-Newton}
  observations (Kotov \& Vikhlinin), we obtained the mass and
  temperature measurements for 13 galaxy clusters at $0.4<z<0.7$
  spanning a temperature interval of $3\,\text{keV}<T<14\,\text{keV}$.
  The observed evolution of the $M-T$ relation, relative to the
  low-redshift references from the \emph{Chandra} sample of Vikhlinin et
  al., follows $M_{500}/T^{3/2} \propto E(z)^{-\alpha}$, where we
  measure $\alpha=1.02\pm0.20$ and $\alpha=1.33\pm0.20$ for the
  spectroscopic and gas mass-weighted temperatures, respectively. Both
  values are in agreement with the expected self-similar evolution,
  $\alpha=1$.  Assuming that the cluster mass for given temperature
  indeed evolves self-similarly, the derived slopes, $\gamma$, of the
  high-redshift $M-T$ relation, $E(z) M_{500} \propto T ^ \gamma$, are
  $\gamma=1.55\pm 0.14$ for $T_{spec}$ and $\gamma=1.65\pm 0.15$ for
  $T_{mg}$. Our results show that both the shape and evolution of the
  cluster $M-T$ relation at $z\simeq 0.5$ is close to predictions of the
  self-similar theory.

\end{abstract}

\keywords{galaxies: clusters: general --- surveys --- X-rays: galaxies}

\section{Introduction}

Scaling relations between the cluster parameters such as total mass,
average gas temperature, and X-ray luminosity can be used for studying
galaxy clusters and their cosmological applications (see Voit 2005 for a
recent review). Several recent \emph{Chandra} and \emph{XMM-Newton}
studies provide accurate measurements for the $M-T$ relation in
low-redshift clusters \citep{Vikhlinin_mp,2005astro.ph..2210A}. While
studying the scaling relations for low-redshift clusters allows to
address many interesting questions, the knowledge of the evolution of
the scaling relations at $z>0$ is required in many cosmological
applications. The X-ray cluster mass measurements based on application
of the hydrostatic equilibrium equation (Mathews 1978; Sarazin 1988)
rely on observations of the cluster temperature and brightness profiles
at large radii. These are technically challenging observations,
especially for distant clusters. In the past, the mass measurements for
distant clusters were often derived assuming $T(r)=\mathrm{const}$
(e.g., Ettori et al. 2004; Maughan et al. 2005).  However, this
assumption can significantly bias the mass measurements
\citep{1997ApJ...491..467M,maughan05}.


The first systematic study of the high-redshift cluster $M-T$ relation
based on the spatially-resolved temperature profiles to $r\approx
r_{500}$ ($r_{500}$ is defined as a radius corresponding mean
overdensity $\Delta=500$ relative to the critical density at a cluster
redshift) from the \emph{XMM-Newton} observations was presented in
\cite{astro-ph/0504233}. In this \emph{Paper}, we expand this work by
analysing \emph{Chandra} observations for 6 clusters at $z=0.4-0.54$.
All our clusters belong to a sample of luminous, dynamically relaxed
systems used by \cite{2004MNRAS.353..457A} for $f_{gas}$ measurements.
Statistical accuracy of the observations from our sample allows us to
trace the surface brightness and temperature profiles nearly to
$r_{500}$.  \emph{Chandra}'s superb angular resolution minimizes the
systematic uncertainties in the spatially-resolved analysis. We use the
derived gas density and temperature profile measurements to infer the
total cluster masses and the average temperatures. A combination of the
\emph{Chandra} and \emph{XMM-Newton} distant cluster samples provides an
accurate determination of the shape and evolution of the cluster $M-T$
relation at $z\simeq 0.5$.

All distance-dependent quantities are derived assuming the
$\Omega_M=0.3$, $\Omega_\Lambda=0.7$ cosmology with the Hubble constant
$H_0=72$km~s$^{-1}$~Mpc$^{-1}$. Statistical uncertainties are quoted at
$68\%$ CL. 

\section{Observations and Data Reduction}

In present analysis, we used the archived \emph{Chandra} observations of
$z>0.4$ clusters. We selected only those objected with the
\emph{Chandra} exposures sufficient for temperature measurements to $r
\sim r_{500}$. We also selected clusters which appear dynamically
relaxed in their {\em Chandra} images.  The selected clusters are listed
in Table\ref{tab:all}. 

Our {\em Chandra} data reduction follows the procedure described in
\cite{2005ApJ...628..655V,Vikhlinin_mp}. This involves generation of the
spatially dependent effective area and detector response files, modeling
of the particle-induced detector background following
\cite{2003ApJ...583...70M}, and subtraction of the residual Galactic
foreground as described in \cite{2005ApJ...628..655V}.


\begin{deluxetable*}{lcccccccccccccc}
\tabletypesize{\scriptsize}
\tablecolumns{15}
\tablewidth{0pc}
\tablecaption{Best Fit Parameters for Gas Density Profiles and Results of Spectral and Mass Determination \label{tab:all}}
\tablehead{
\colhead{Cluster\hspace*{17mm}} &
\colhead{$z$} &
\colhead{$n_{02}$/$n_0$\notea} &
\colhead{$r_c$\notea} &
\colhead{$r_s$\notea} &
\colhead{$\alpha$\notea} &
\colhead{$\beta$\notea} &
\colhead{$\varepsilon$\notea} &
\colhead{$r_{c2}$\notea} &
\colhead{$\beta_2$\notea} &
\colhead{$r_{500}$} &
\colhead{$T_{\text{spec}}$} &
\colhead{$T_{\text{mg}}$} &
\colhead{$M_{500}$}
\\
 &  &  & \colhead{(kpc)} & \colhead{(kpc)} &  &  &  &\colhead{(kpc)}  & &
\colhead{(kpc)}&\colhead{(keV)}&\colhead{(keV)} & \colhead{($10^{14}\,M_\odot$)}}
\startdata
MACS1423.8+2404\dotfill & 0.539&   14.9 &    28 &   1220 &  1.86 &  0.53 &  4.94 &   0.02 &   1.00 & $ 973\pm72$ & $7.02\pm0.28$ & $6.27\pm0.40$ & $ 4.56\pm1.04$\\
3C295         \dotfill & 0.460&  173.5 &   166 &   1548 &  2.47 &  0.63 &  4.99 &   5.80 &   1.00 & $ 840\pm66$ & $5.13\pm0.24$ & $4.52\pm0.34$ & $ 2.63\pm0.66$\\
MACSJ0159.8-0849\dotfill & 0.405&   27.3 &   247 &   2308 &  1.84 &  0.69 &  5.00 &  23.74 &   1.00 & $1324\pm127$ & $9.59\pm0.50$ & $9.62\pm1.01$ & $ 9.86\pm3.12$\\
RXJ1347.5-1145\dotfill & 0.451&   13.0 &   116 &   1821 &  2.17 &  0.60 &  5.00 &   5.93 &   3.77 & $1446\pm100$ & $14.03\pm0.69$ & $12.05\pm0.82$ & $13.83\pm2.82$\\
MACSJ1621.3+3810\dotfill & 0.461&   34.7 &   263 &   2059 &  2.29 &  0.67 &  4.96 &  16.73 &   1.00 & $ 992\pm71$ & $7.53\pm0.41$ & $6.42\pm0.48$ & $ 4.37\pm1.01$\\
MACSJ0329.6-0211\dotfill & 0.450&   26.6 &    50 &    466 &  2.14 &  0.49 &  1.59 &   0.00 &   1.00 & $ 915\pm83$ & $5.24\pm0.38$ & $5.23\pm0.60$ & $ 3.64\pm1.02$
\enddata
\tablenotetext{a}{~The obtained best-fit parameters of the 3D gas density distribution. For definition  see Eq.3 in \cite{Vikhlinin_mp}}
\end{deluxetable*}

\begin{figure}
\centerline{\includegraphics[width=1.0\linewidth]{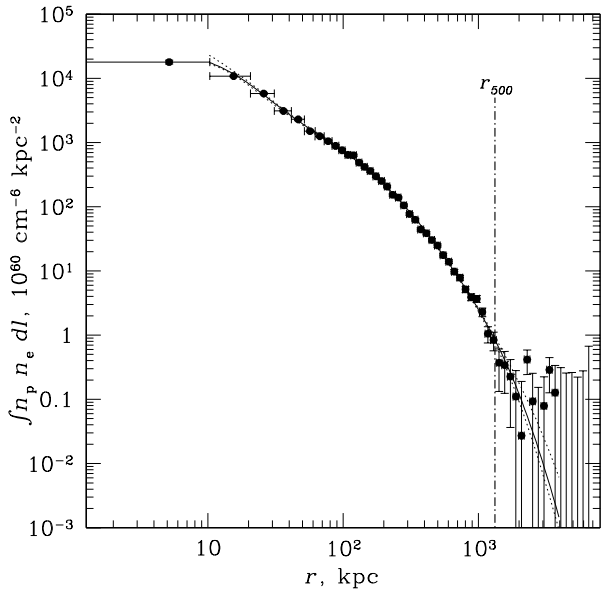}}
\caption{Observed emissivity profile for MACSJ0159.8-0849. The solid line shows the best-fit  profile and the dashed lines correspond to its $68\%$  CL uncertainties} 
\label{fig:em}
\end{figure}
\begin{figure*}
\centerline{%
\includegraphics[width=0.33\linewidth]{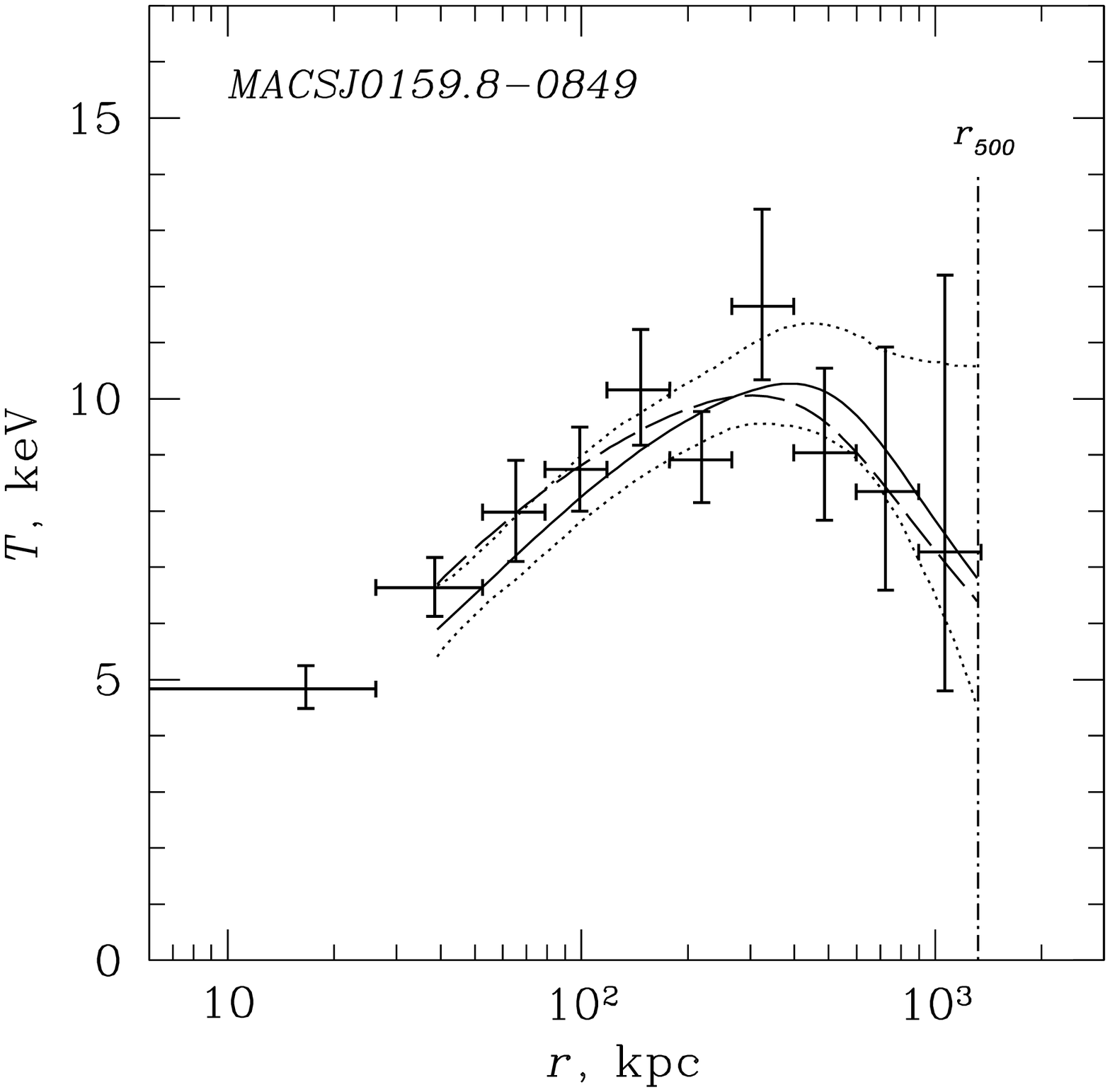}~~~%
\includegraphics[width=0.33\linewidth]{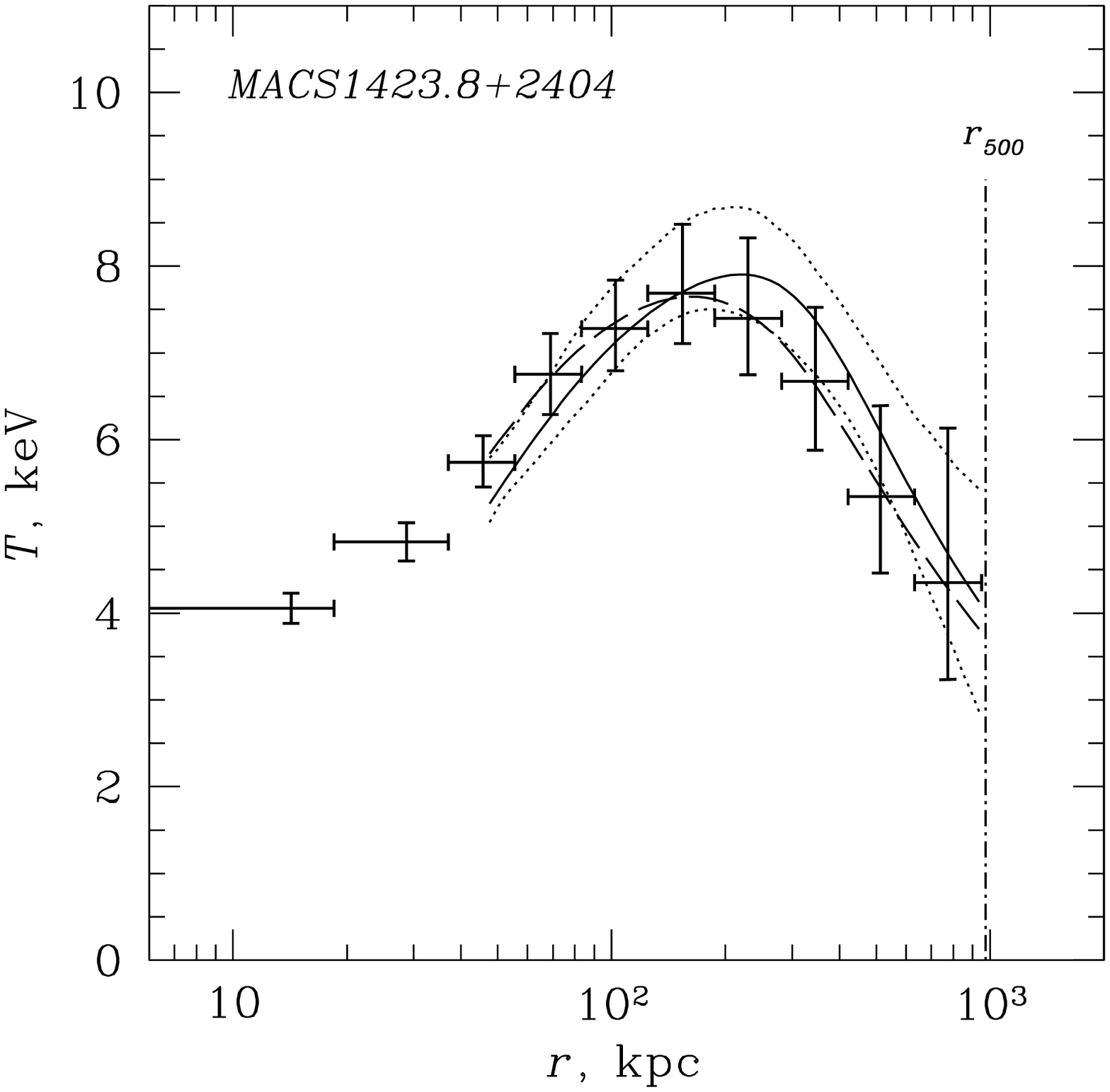}~~~%
\includegraphics[width=0.33\linewidth]{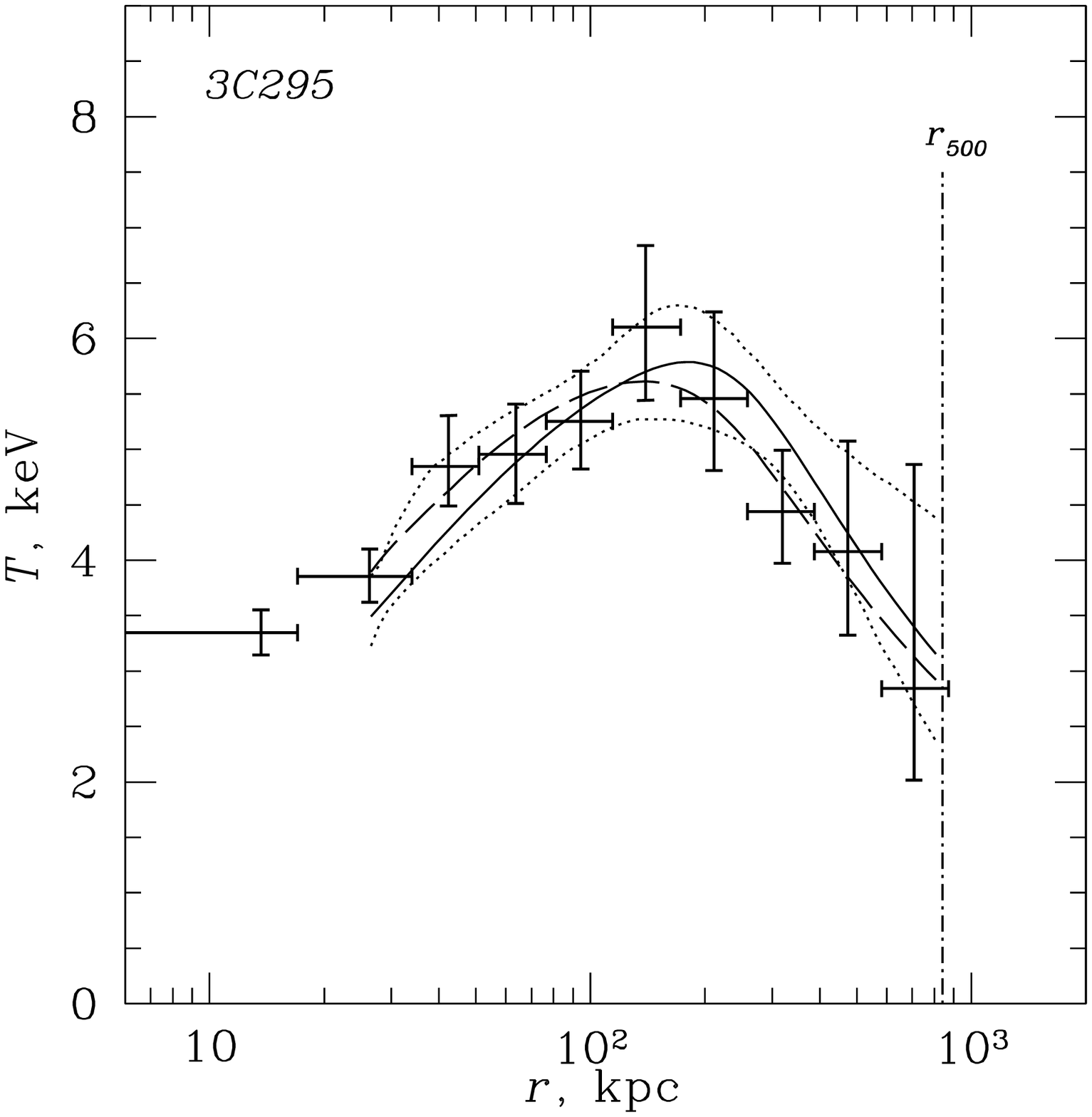}
}
\medskip
\centerline{%
\includegraphics[width=0.33\linewidth]{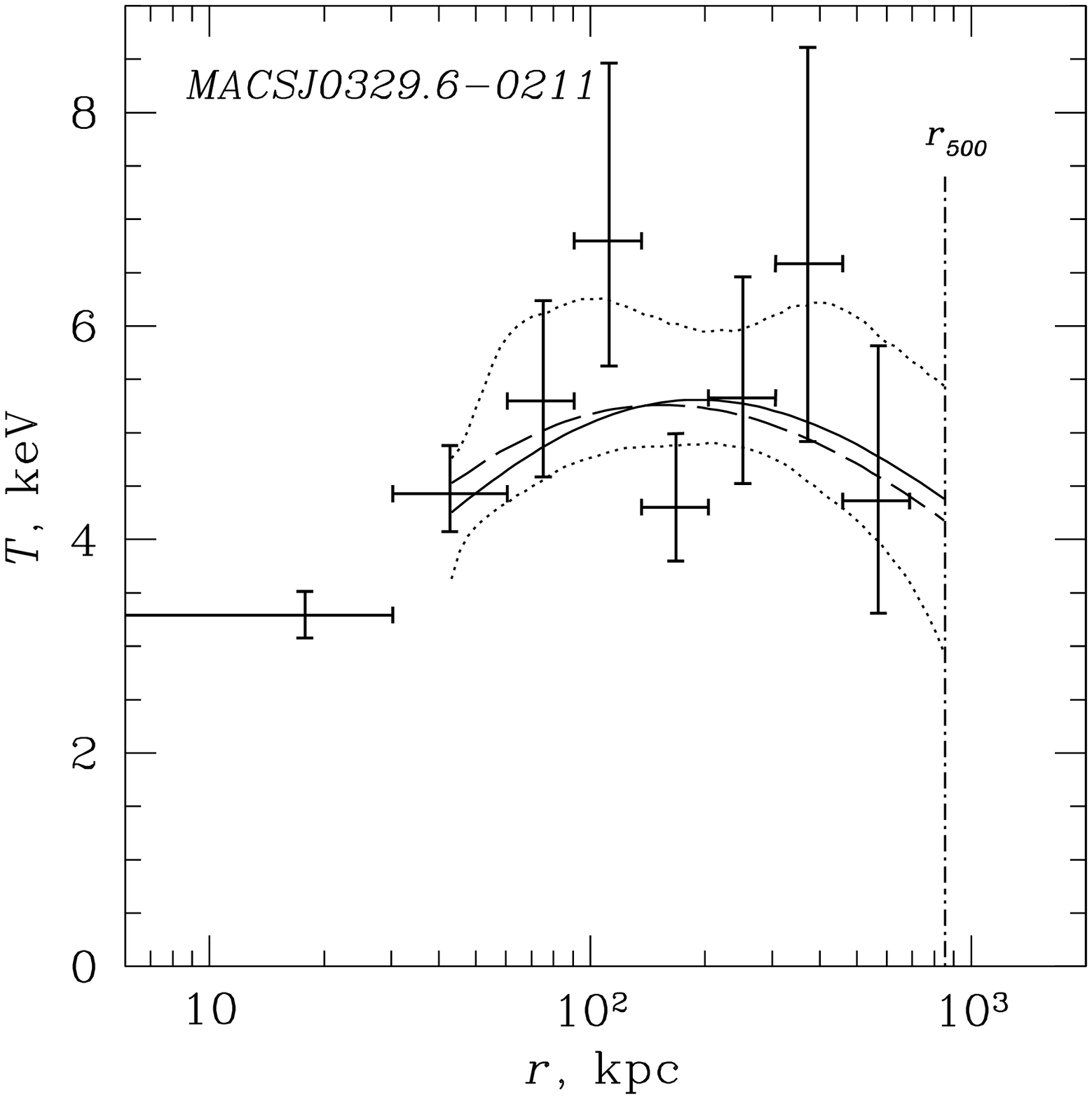}~~~%
\includegraphics[width=0.33\linewidth]{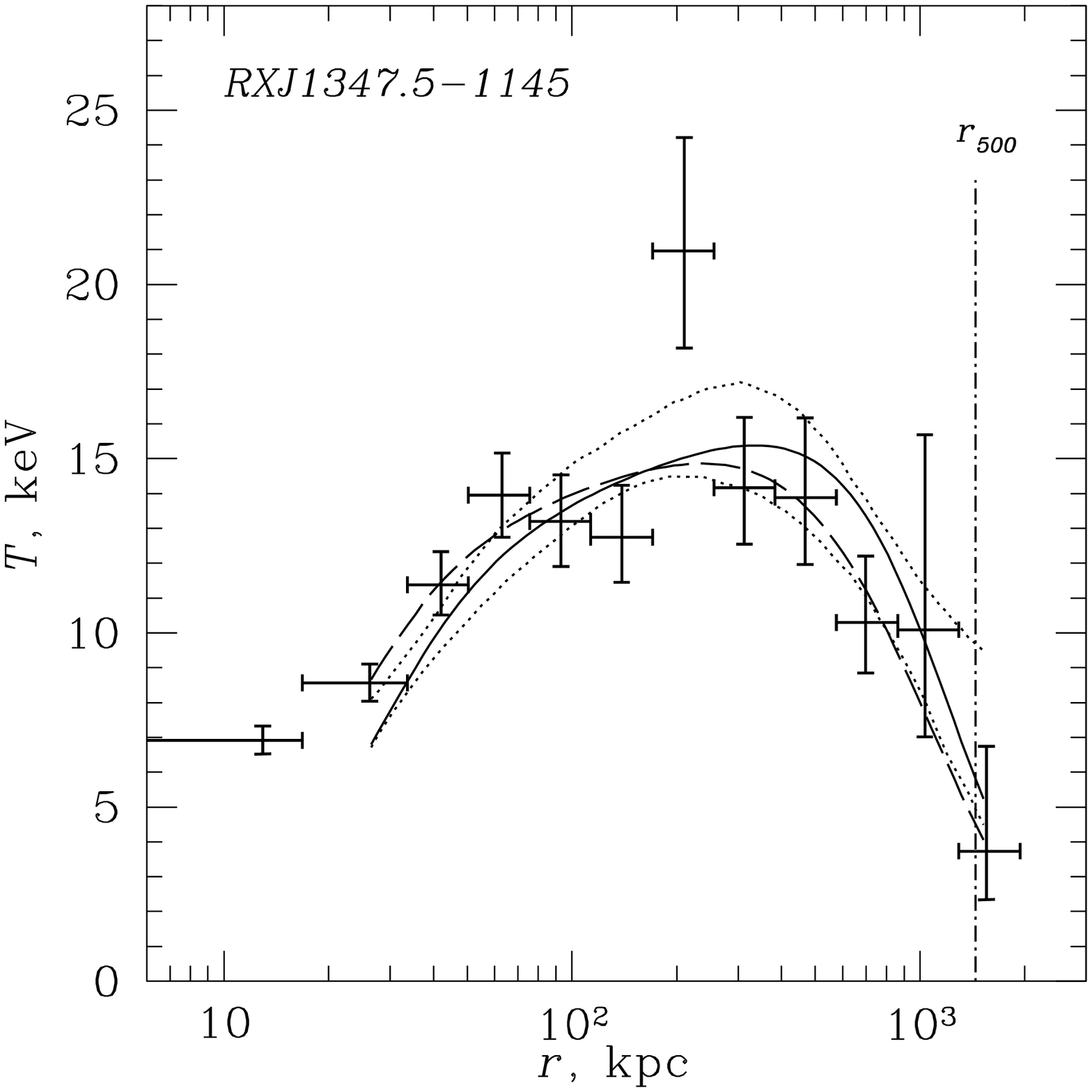}~~~%
\includegraphics[width=0.33\linewidth]{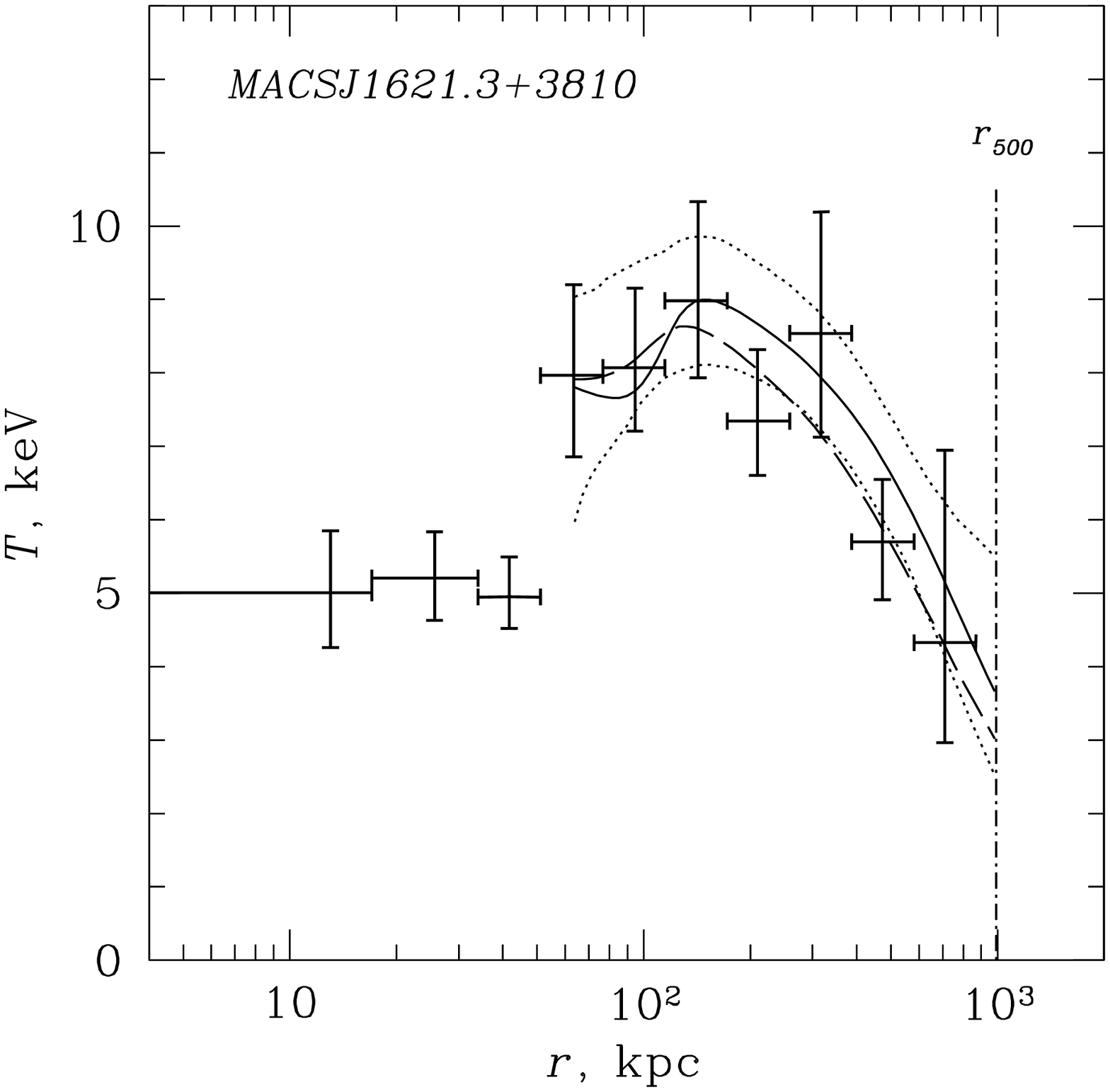}
}
\caption{Observed temperature profiles of a sample of distant clusters
  observed with {\em Chandra}. The solid lines show the best-fit 3-D
  temperature profiles and the dotted lines correspond to their $68\%$
  CL uncertainties. The dashed lines show the projected best-fit 3-D
  temperature profiles (2-D). For MACSJ0329.6-0211, we do not show the
  two outermost temperature bins because they can be contaminated by the
  foreground emission (see text); these bins were not used in the 3D
  modeling.}
\label{fig:tprof}
\end{figure*}

All detectable point and small-scale extended sources were masked out
from the analysis of azimuthally averaged surface brightness and
temperature profiles. The sources were detected using a wavelet
decomposition technique described in \cite{1998ApJ...502..558V}. The
sources were automatically detected in the 0.7--2~keV and 2--7~keV
energy bands and the merged region mask was checked manually.

All visible large-scale extended X-ray sources were also excluded from
the data.  Such structures are present in 3 clusters.  In
MACSJ0159.8-0849 there are detectable filamentary structures with a size
of $\sim 2'$, located at $\simeq 5\arcmin$ from the cluster
center (at ${\rm RA} = 30.031\arcdeg$; ${\rm Dec} = -8.854\arcdeg$ and
${\rm RA} = 29.965\arcdeg$; ${\rm Dec} = -8.921\arcdeg$).
MACSJ1621.3+3810 is projected on the outskirts of a low-redshift galaxy
group ($\sim25\arcmin$ from the group center). The X-ray emission
associated with the group is clearly detected in the Southern part of
the \emph{Chandra} image and is undetectable in the Northern part.  In
this case, we excluded the data within the lower (Southern) 50\%
of the field of view. We also excluded from the analysis the two
outermost bins in the MACSJ1621.3+3810 temperature profile because of
potential contamination by the foreground emission. In addition, there
is a small-scale extended source at ${\rm RA} = 245.296\arcdeg$; ${\rm
  Dec} = 38.209\arcdeg$, also masked out.  The only cluster in our
sample that shows some signs of a recent merger is RXJ1347.5--1145
\citep{2002MNRAS.335..256A,2004A&A...427L...9G}. In this case, we masked
out the X-ray subclump at $20''$ Southeast of the cluster center. We
also excluded the sector $\mathrm{PA}=180\arcdeg-225\arcdeg$. This
sector shows an excess of the X-ray surface brightness compared to all
other directions, which is probably associated with the trail of the
merging subcluster.

The three-dimensional gas density and temperature profiles were derived
from the azimuthally averaged X-ray brightness and projected temperature
profiles centered on the cluster X-ray brightness peak.  
The projected temperatures were measured using the spectra accumulated
in the concentric annuli with $r_{out}/r_{in}=1.5$. 
These spectra were fit to the single-temperature MEKAL model
\citep{1985A&AS...62..197M} with Galactic absorption. The absorber's
hydrogen column density was fixed at values provided by HI radio surveys
\citep{1990ARA&A..28..215D}. The metallicity was allowed to be free in
almost all annuli except the outermost ones, where it was fixed at an
average value from the two outermost bins where direct measurements were
possible.

The X-ray brightness profiles were extracted from the images in the
0.7--2~keV energy band in concentric annuli with $r_{out}/r_{in}=1.1$. 
Using the observed projected temperature and metallicity profiles, the
raw \emph{Chandra} counts were converted to the profiles of the
projected emission measure, $\int n_e n_p d l$, as described in
\cite{Vikhlinin_mp}. The profiles from different pointings were added
using the statistically optimal weighting. The resulting EM profiles
were used to derive the gas density profile.

\section{Profile modeling and determination of the cluster mass}

The 3-dimensional gas density profile was derived by fitting the
observed emission measure profile to a model used by \cite{Vikhlinin_mp}
to describe the 3-D gas density distribution in the low-redshift
\emph{Chandra} sample (see their equation 3). This function is a
modification of the so-called $\beta$-model \citep{1978A&A....70..677C},
designed to independently fit the gas density slopes in the cluster
center, outskirts, and in the intermediate region, and also to allow for
an additional emission component in the very central region. The
statistical quality of the \emph{Chandra} data for high-$z$ clusters (an
example is shown in Fig.\ref{fig:em}) is sufficient to fit the same set
of parameters that \cite{Vikhlinin_mp} used for the low-$z$ clusters. 
The best-fit parameters of the gas density profiles are listed in Table
\ref{tab:all}. The projected emission measure corresponding to the
best-fit model in MACSJ0159.8-0849 is shown by solid line in
Fig.\ref{fig:em}. 

For derivation of the 3-dimensional gas temperature profile we also
followed the procedure in \cite{Vikhlinin_mp}. Specifically, a model
with a great functional freedom was assumed for the 3D profile
\citep[see equation 6 in][]{Vikhlinin_mp}. It was projected along the line
of sight using the best-fit gas density profile. The projection uses the
method that accurately predicts the single-temperature fit for a mixture
of plasma spectra with different $T$
\citep{Vikhlinin_w,2004MNRAS.354...10M}. The parameters of the 3D model
were obtained by minimizing $\chi2$ computed using the projected model
and observed temperature values at each radius. In some cases, the
functional freedom of the 3D model was insufficient to accurately
describe the data simultaneously in the very central region and at large
radii. In such cases, we excluded several innermost temperature
measurements since our prime goal is the mass determination at large
radii.  The observed temperature profiles are shown along with the best
fit models in Fig.\ref{fig:tprof}. 

The total cluster mass profile was derived by direct application of the
hydrostatic equilibrium equation (e.g., Sarazin 1988) to the best fit
gas density and temperature profiles. The mass profile was used to find
the radius corresponding to the mean spherical overdensity $\Delta=500$
relative to the critical density at the cluster redshifts. The
corresponding masses, $M_{500}$, are reported in Table \ref{tab:all} and
the values of $r_{500}$ are indicated in Fig.\ref{fig:em} and
\ref{fig:tprof}.

Using the best-fit three dimensional profiles, we also computed the
average temperatures excluding the central region, as was done in
\cite{Vikhlinin_mp} for the low-redshift clusters. We derived the
average spectroscopic temperature, $T_{spec}$, 
a value obtained from a single-temperature fit  to the integrated  spectrum, 
and computed the gas  mass-weighted average $T_{mg}$, both in the
70~kpc--$r_{500}$ radial range. These averages were used by
\cite{Vikhlinin_mp} to derive the $M-T$ relations for nearby clusters
and therefore we can use their results as the low-redshift reference.

All the essential parameters derived from the 3-dimensional modeling are
reported in Table~\ref{tab:all}. The parameter uncertainties were
estimated by the Monte-Carlo simulations technique described in \S~3.3
and 3.4 of \cite{Vikhlinin_mp}. 

\begin{deluxetable}{lcc}
\tablecaption{Power Law Fit to Combined {\em Chandra} and {\em XMM-Newton}
Mass-Temperature  Relation \label{tab:mt}}
\tablehead{
\colhead{Temperature average} &
\colhead{$\alpha$} &
\colhead{$\gamma$}
}
\startdata
$T_{spec}$\dotfill&  $1.02\pm0.20$    &  $1.55\pm 0.14$    \\
$T_{mg}$\dotfill&    $1.33\pm0.20$    &   $\,\,\,\,1.65\pm 0.15$
\enddata
\tablecomments{Parameters $\gamma$ characterize the slope of the $M-T$
  relation, $E(z) M_{500} \propto T^\gamma$, and evolution in its
  normalization, $M_{500}/T^{3/2}\propto E(z)^{-\alpha}$. See text for
  details.}
\end{deluxetable}

\begin{figure}
\centerline{\includegraphics[width=1.0\linewidth]{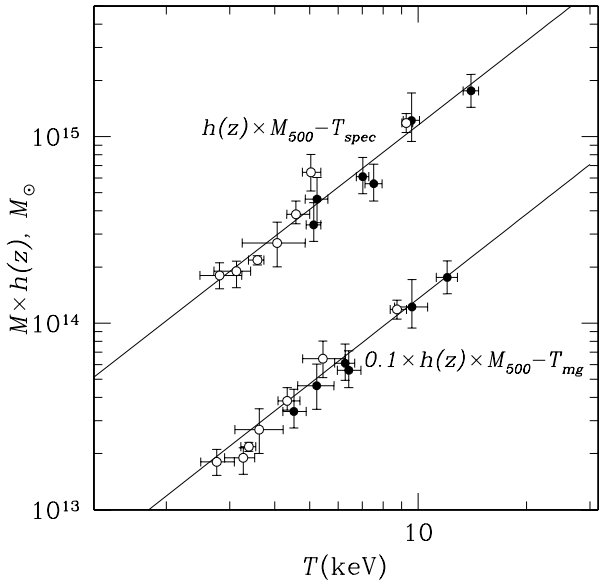}}
\caption{Correlation of $M-T_{spec}$ (the upper-left corner) and $M-T_{mg}$ (the lower-right corner) at $r_{500}$ for the
combined {\em XMM-Newton}(open circles) and  {\em Chandra}(solid circles) sample.  
The solid lines show the low-redshift results from \cite{Vikhlinin_mp}. 
} 
\label{fig:xmm_m_t}
\end{figure}

\section{$M-T$ relation}
\label{sec:mt_relation}
Figure~\ref{fig:xmm_m_t} shows the derived $M-T$ relation for $z>0.4$,
which combines the \emph{XMM-Newton} results from
\cite{astro-ph/0504233} and the \emph{Chandra} results described here. 
There is clearly a good overall agreement between the two datasets. 
There is a small difference in our modeling procedure for the
\emph{XMM-Newton} and \emph{Chandra} data (we used more restrictive
models for $\rho_g(r)$ and $T(r)$ to fit the \emph{XMM-Newton} data). 
However, we checked that this difference results in a negligibly small
offset in the mass estimates. Applying the more restrictive models that
were used in the \emph{XMM-Newton} analysis to the \emph{Chandra} data,
we obtain that the derived values of $M_{500}$ change by $3\%$ on
average.

To quantify the high-redshift $M-T$ relation, we model separately the
evolution in normalization and the slope of de-evolved relation. The
evolution in the normalization is parametrized as $M_{500}/T^{3/2}
\propto E(z)^{-\alpha}$, where the low-redshift normalization is fixed at
the \emph{Chandra} measurements by \cite{Vikhlinin_mp}. The joint fit to
the \emph{Chandra} and \emph{XMM-Newton} results for distant clusters
gives $\alpha=1.02\pm0.20$ and $\alpha=1.33\pm0.20$ for the $M-T_{spec}$
and $M-T_{mg}$ relations, respectively\footnote{Uncertainties include
  that for the normalization of the low-$z$ mass-temperature relations
  from \cite{Vikhlinin_mp}.}  (see also
Table~\ref{tab:mt}).  These values are close to, and consistent with,
the expected self-similar evolution, $\alpha=1$. 

Assuming that the normalization of the $M-T$ relation indeed evolves
self-similarly, we can ``de-evolve'' each mass measurement using the
correction factor $E(z)$ and fit the slope in the high-redshift $M-T$
relation, $E(z) M_{500} \propto T^\gamma$. For the $M-T_{spec}$ and
$M-T_{mg}$ relations we measure slopes of $\gamma=1.55\pm 0.14$ and
$\gamma=1.65\pm 0.15$, respectively ( see
also Table~\ref{tab:mt}).  These values are also close to the
predictions of the self-similar theory, $\gamma=3/2$, and to the
low-redshift measurements \citep{Vikhlinin_mp,2005astro.ph..2210A}.

\section{Conclusions}

Deep \emph{Chandra} and \emph{XMM-Newton} observations of relaxed
clusters at $z>0.4$ provide direct hydrostatic mass measurements at the
critical overdensity level $\Delta=500$. The resulting determination of
the high-redshift $M-T$ relation is comparable to the low-redshift studies
both in terms of the sample size and mass measurement uncertainties for
individual objects. 

We find that the $M-T$ relation at $z=0.4-0.7$ is consistent with the
self-similar evolution of that for low-redshift clusters. Parametrizing
the evolution of the normalization as $M/T^{3/2}\propto E(z)^{-\alpha}$
we measure $\alpha$ close to 1 (Table.~\ref{tab:mt}). The de-evolved
relation, $E(z) M-T$, has the power law slopes close to $\gamma=1.5$
(Table~\ref{tab:mt}). This is an important confirmation of one of the
key predictions of the cluster formation theory
\citep{1996MNRAS.282..263E,1998ApJ...495...80B} and also a welcome news
for cosmological interpretation of the cluster number density
measurements at $z>0$. 

We should caution, however, that our results are based on the data for
highly relaxed clusters, which represent only a fraction of the total
population, especially at high redshifts. Extrapolation of the observed
$M-T$ relation on the entire population requires further studies with
the help of numerical simulations and independent mass measurements from
gravitational lensing. 

\acknowledgements

This work was supported by NASA grant NAG5-9217 and contract NAS8-39073. 
O. K. thanks SAO for hospitality during the course of this research.  We
thank S.~Allen for providing redshifts for several MACS clusters.


\begin{thebibliography}{}


\bibitem[Allen et al.(2002)]{2002MNRAS.335..256A} Allen, S.~W., Schmidt, 
R.~W., \& Fabian, A.~C.\ 2002, \mnras, 335, 256 
\bibitem[Allen et al.(2004)]{2004MNRAS.353..457A} Allen, S.~W., Schmidt, 
R.~W., Ebeling, H., Fabian, A.~C., \& van Speybroeck, L.\ 2004, \mnras,  353, 45
\bibitem[Arnaud et al.(2005)]{2005astro.ph..2210A} Arnaud, M., 
Pointecouteau, E., \& Pratt, G.~W.\ 2005, \aap, 441, 893 
\bibitem[Bryan \& Norman(1998)]{1998ApJ...495...80B} Bryan, G.~L., \& 
Norman, M.~L.\ 1998, \apj, 495, 80 
\bibitem[Cavaliere \& Fusco-Femiano(1978)]{1978A&A....70..677C} Cavaliere, 
A., \& Fusco-Femiano, R.\ 1978, \aap, 70, 677 
\bibitem[Dickey \& Lockman(1990)]{1990ARA&A..28..215D} Dickey, J.~M., \& 
Lockman, F.~J.\ 1990, \araa, 28, 215 
\bibitem[Eke et al.(1996)]{1996MNRAS.282..263E} Eke, V.~R., Cole, S., \& 
Frenk, C.~S.\ 1996, \mnras, 282, 263 
\bibitem[Ettori et al.(2004)]{2004A&A...417...13E} Ettori, S., Tozzi, P., 
Borgani, S., \& Rosati, P.\ 2004, \aap, 417, 13 
\bibitem[Gitti \& Schindler(2004)]{2004A&A...427L...9G} Gitti, M., \& 
Schindler, S.\ 2004, \aap, 427, L9 
\bibitem[Kotov \& Vikhlinin(2005)]{astro-ph/0504233} Kotov, O., \&
Vikhlinin, A.\ 2005, ApJ, in press,  (astro-ph/0504233)
\bibitem[Markevitch \& Vikhlinin(1997)]{1997ApJ...491..467M} Markevitch, 
M., \& Vikhlinin, A.\ 1997, \apj, 491, 467 
\bibitem[Markevitch et al.(2003)]{2003ApJ...583...70M} Markevitch, M., et 
al.\ 2003, \apj, 583, 70 
\bibitem[Mathews(1978)]{1978ApJ...219..413M} Mathews, W.~G.\ 1978, \apj, 
219, 413 
\bibitem[Maughan et al.(2005)]{maughan05} Maughan, B.~J., Jones, L.~R.,
  Ebeling, H., \& Scharf, C.\ 2005, mnras, accepted for publication (astro-ph/0503455)
\bibitem[Mazzotta et al.(2004)]{2004MNRAS.354...10M} Mazzotta, P., Rasia, 
E., Moscardini, L., \& Tormen, G.\ 2004, \mnras, 354, 10 
\bibitem[Mewe et al.(1985)]{1985A&AS...62..197M} Mewe, R., Gronenschild, 
E.~H.~B.~M., \& van den Oord, G.~H.~J.\ 1985, \aaps, 62, 197 
\bibitem[Sarazin(1988)]{1988xrec.book.....S} Sarazin, C.~L.\ 1988, 
Cambridge Astrophysics Series, Cambridge: Cambridge University Press, 1988, 
\bibitem[Vikhlinin et al.(1998)]{1998ApJ...502..558V} Vikhlinin, A., 
McNamara, B.~R., Forman, W., Jones, C., Quintana, H., \& Hornstrup, A.\ 1998, \apj, 502, 558
\bibitem[Vikhlinin et al.(2005a)]{2005ApJ...628..655V} Vikhlinin, A., 
Markevitch, M., Murray, S.~S., Jones, C., Forman, W., \& Van Speybroeck, 
L.\ 2005a, \apj, 628, 655 
\bibitem[Vikhlinin et al.(2005b)]{Vikhlinin_mp}Vikhlinin, A., Kravtsov A., Forman, W., Jones, C., Markevitch, M., Murray, S.~S., \& Van Speybroeck \ 2005b, ApJ submitted, (astro-ph/0507092) 
\bibitem[Vikhlinin(2005)]{Vikhlinin_w}Vikhlinin, A., 2005, ApJ, submitted, (astro-ph/0504098)
\bibitem[Voit(2005)]{voit2005} Voit, G.~M.\ 2005, Reviews of Modern Physics, 77, 207 
\end{thebibliography}
\end{document}